\input epsf
\def\tightitemize{\begin{itemize}\setlength{\parskip}{0.5 \parsep}
                  \setlength{\itemsep}{0 pt}}

\def\endcvitemize{\end{itemize} \end{itemize} \end{itemize}}
\def\selectedbib{\section[Selected Bibliography]{Selected Bibliography}
\bgroup\parindent=0pt\parskip=\itemsep
\def\refpar{\par\hangindent=3em\hangafter=1}}
\def\endselectedbib{\refpar\egroup}

\def\agt{>\kern-1.1em \lower1.1ex\hbox{$\sim$}\kern.3em}
\def\alt{<\kern-1.1em \lower1.1ex\hbox{$\sim$}\kern.3em}
\def\3he{$^3{\rm He}$}

\def\gs{\mathrel{\raise0.35ex\hbox{$\scriptstyle >$}\kern-0.6em 
\lower0.40ex\hbox{{$\scriptstyle \sim$}}}}
\def\ls{\mathrel{\raise0.35ex\hbox{$\scriptstyle <$}\kern-0.6em 
\lower0.40ex\hbox{{$\scriptstyle \sim$}}}}
\newfont{\rmsmall}{cmr12 scaled 900}

\def\litl{\rm\scriptscriptstyle}

\def\carbon#1{\ifmmode{{^{#1}{\rm C}}}\else{{$^{#1}{\rm C}$}}\fi}

\def\nnh{\ifmmode{{{\rm N}_{\litl H}}}\else{{${\rm N}_{\litl H}$}}\fi}

\def\nc{\ifmmode{{{\rm N}_{\litl C}}}\else{{${\rm N}_{\litl C}$}}\fi}
\def\nco{\ifmmode{{{\rm N}_{\litl CO}}}\else{{${\rm N}_{\litl CO}$}}\fi}

\def\car#1 {\alwaysmath{{}^{#1}{\rm C}}}

\documentstyle[aaspp4, 12pt, hacks, myreferences]{article}
\renewcommand{\thesubsection}{\arabic{subsection}}

\begin{document}
\title{Meeting the Optical Requirements of\\
Large Focal-Plane Arrays}
\vskip 16pt
\noindent
\author{${\rm Antony~A.~Stark}$ \\
Smithsonian Astrophysical Observatory
}
\textwidth=6.0in
\vskip 20pt
\begin{abstract}{
\noindent
Technological advances will allow the placement
of many Terahertz detectors at the 
focal plane of a single telescope.
For a telescope of a given diameter and wavelength
of operation, there is a limit to the
number of usable detectors 
imposed by diffraction and optical aberrations.
These effects can be ameliorated through an optical
design where the magnification of the 
telescope's secondary mirror is small 
and the detector package is therefore located near
the secondary mirror.
A field mirror can be used to flatten the image,
and the focal reducer which matches the detector to
the telescope can also provide an image of the
aperture for placement of filters and stops.
A design concept is presented for the South Pole
Telescope which comprises a 10 meter diameter 
off-axis paraboloidal primary mirror, a Gregorian
secondary mirror, a tertiary chopper,  
dewar widow, Lyot stops,
band-pass filter, and space
behind the focal plane for cryogenics.
The telescope is bilaterally symmetric, and
all apertures are unblocked.
The field of view is one degree in diameter,
so this telescope can feed an 
array of several $\times \, 10^4$
detectors at Terahertz frequencies. 

}
\end{abstract}

\textwidth=6.5in

\section{Introduction}

The South Pole Telescope (SPT) is a 10 m diameter millimeter-
and submillimeter-wave telescope which has been approved for construction
at Amundsen-Scott South Pole Station.  It is funded by the
National Science Foundation Office of Polar Programs, and
the participating institutions are
the University of Chicago, the Smithsonian Astrophysical Observatory,
the University of California at Berkeley,
Case Western Reserve University and the University of Illinois.
Work on the telescope design is in progress.
Construction is expected to begin by the end of 2003.

The initial project planned for this telescope is a large-scale survey
covering many thousands of square degrees of sky, searching for
the Sunyaev-Zel'dovich effect from clusters of galaxies at all redshifts.
This search will be done with a focal-plane array of bolometers
operating at 2 mm wavelength.  
The diffraction-limited beamsize at that wavelength (with conservative
illumination of the primary mirror) will be about an arcminute, an angular
size chosen to be comparable to the size of clusters at
cosmological distances.  The focal plane array will have about 1000 
detectors; making an array of this size is thought to be possible using
current bolometer technology 
(\cite{gildemeister00}).
Such a large number of independent beams of arcminute size requires
a field of view about a degree in diameter.
This paper describes a design for the telescope optics which will
feed 1000 beams at $\lambda = 2 \mathrm{mm}$. 
The significance of this work for Terahertz technology is the
serendipitously good quality to which this design can be optimized:
the optics are
sufficiently well-corrected
that they will work at
$\lambda = 200 \mu \mathrm{m}$ and feed an array
containing more than 30,000 detectors.

\section{Optical Design Considerations}

Submillimeter-wave telescopes have different design constraints than
optical telescopes.  This is fortunate, because 10 meter class 
visual-wavelength telescopes have fields of view less than 0.2 degree 
in diameter,
much smaller than the design requirement of the SPT.  Since diffraction is
more important at submillimeter wavelengths, the allowable
size of optical aberrations is larger, and if the aberrations are
balanced across the field of view, that field of view can be large
(\cite{stark00,stark98b}).

The SPT design is driven by its intended use as a cosmic microwave
background instrument. It must have the lowest possible thermal
emission from the telescope, the lowest possible variations in
the residual thermal emission that remain, and the smallest possible
spillover of the beam onto the ground and surrounding structures.
This means that the optical path should have
\begin{itemize}
\item{no aperture blockage anywhere in the system;}
\item{conservatively illuminated optics;}
\item{no warm lenses.}
\end{itemize}
The optical train must therefore consist almost entirely of off-axis
mirrors, with the possible exception of a lens or lenses
which are inside the cryogenic dewar and cooled to low temperatures.
Fortunately, machined aluminum mirrors are near-perfect at
millimeter and submillimeter wavelengths; the addition of a mirror to
the optical train does not cause significant degradation of the system
and many mirrors can be used.
Furthermore, it is possible to fabricate mirrors of arbitrary shape.

\newpage

\noindent
The cryogenic dewar brings its own set of optical requirements:
\begin{itemize}
\item{the dewar window must be small enough that it can be
constructed from low-loss materials and not implode;}
\item{there should be a Lyot stop---an image of the primary mirror inside
the dewar---for placement of a cold aperture which prevents the
detectors from seeing beyond the edges of the optical elements;}
\item{there must be a low-pass filter near the dewar window to
block infrared radiation;}
\item{the filter diameters must be small enough 
to be constructible---also, filters do not work at 
large angles of incidence.}
\end{itemize}
It is best that the dewar not be too large, since the
weight of the dewar increases and the ease of handling
decreases as the enclosed volume goes up.
The Lyot stop should therefore be close to the detector, since
both must be kept cold.  Mirrors can be warm and need not be
inside the dewar.
The detectors also impose their own requirements:
\begin{itemize}
\item{there must be a band-pass filter to determine the 
wavelength of operation;}
\item{the ratio f/D should be a uniform, fixed value (1.3 in this
case) for each point in the image;}
\item{the chief ray at each detector should be perpendicular to
the surface of best images; that surface should be flat;}
\item{all detectors should be identically illuminated.}
\end{itemize}

\begin{figure}[p!]
\begin{center}
\begin{minipage}{3.7in}
\vskip 10 pt
\epsfxsize=2.7in
\ \ \epsfbox{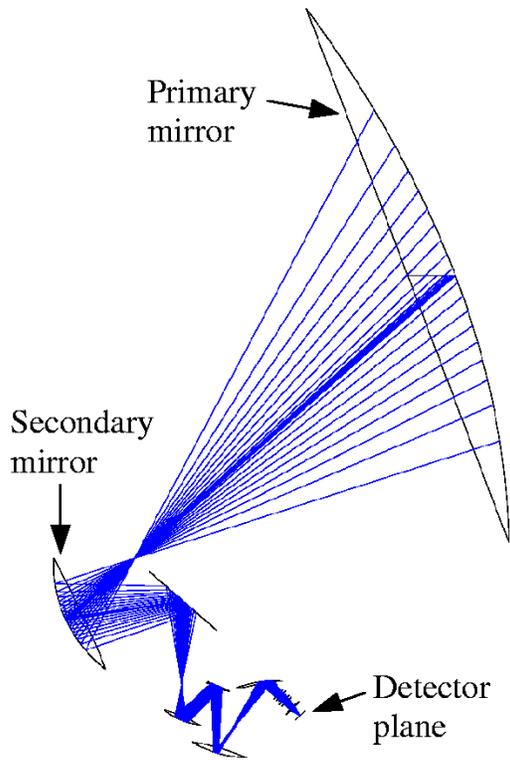}
\end{minipage}
\begin{minipage}{2.4in}
\caption[Side view of SPST214]
{{\bf Side view of SPST214\ \ }
Only rays from the central point in the field are shown.
\label{fig:bigsidef1}
\vfill}
\end{minipage}
\end{center}
\end{figure}

\begin{figure}[p!]
\begin{center}
\begin{minipage}{3.7in}
\vskip 10 pt
\epsfxsize=2.7in
\ \ \epsfbox{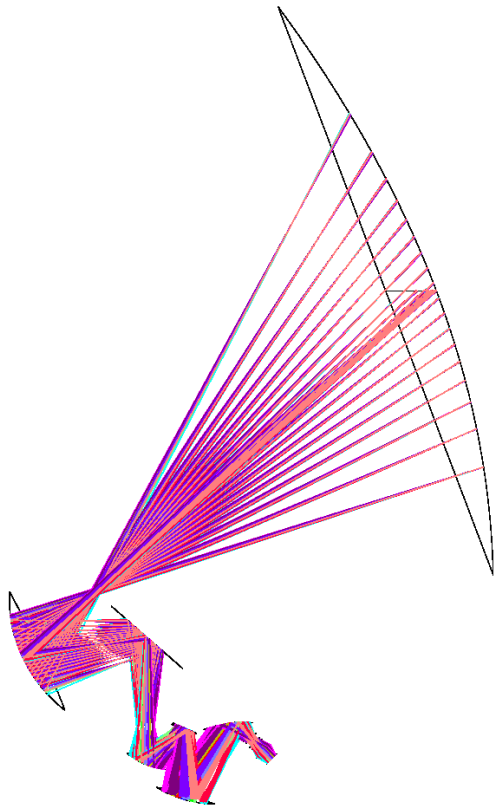}
\end{minipage}
\begin{minipage}{2.4in}
\caption[Side view of SPST214]
{{\bf Side view of SPST214\ \ }
Rays from twelve points in the field of view are shown.
This fills out the volume occupied by the optical beams, 
showing that the fourth through seventh mirrors are neither
blocked nor overilluminated, even in this design with
a large Lagrange invariant.
\label{fig:bigsideall}
\vfill}
\end{minipage}
\end{center}
\end{figure}

\begin{figure}[p!]
\begin{center}
\begin{minipage}{3.5in}
\vskip 10 pt
\epsfxsize=3.5in
\ \ \epsfbox{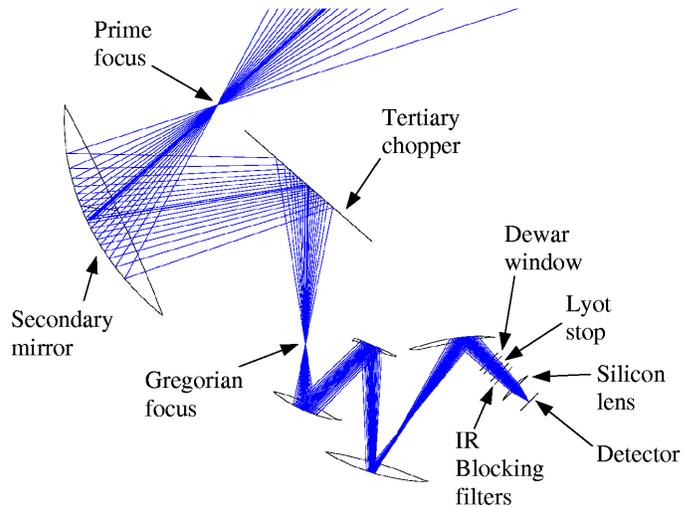}
\vskip 20 pt
\end{minipage}
\begin{minipage}{2.6in}
\caption[Detail view of Figure~\ref{fig:bigsidef1}]
{{\bf Detail view of Figure~\ref{fig:bigsidef1}.}
Only rays from the central point in the field are shown.
\label{fig:smallsidef1}
\vfill}
\vskip 10 pt
\end{minipage}
\end{center}
\end{figure}

\begin{figure}[p!]
\begin{center}
\begin{minipage}{3.5in}
\vskip 10 pt
\epsfxsize=3.5in
\ \ \epsfbox{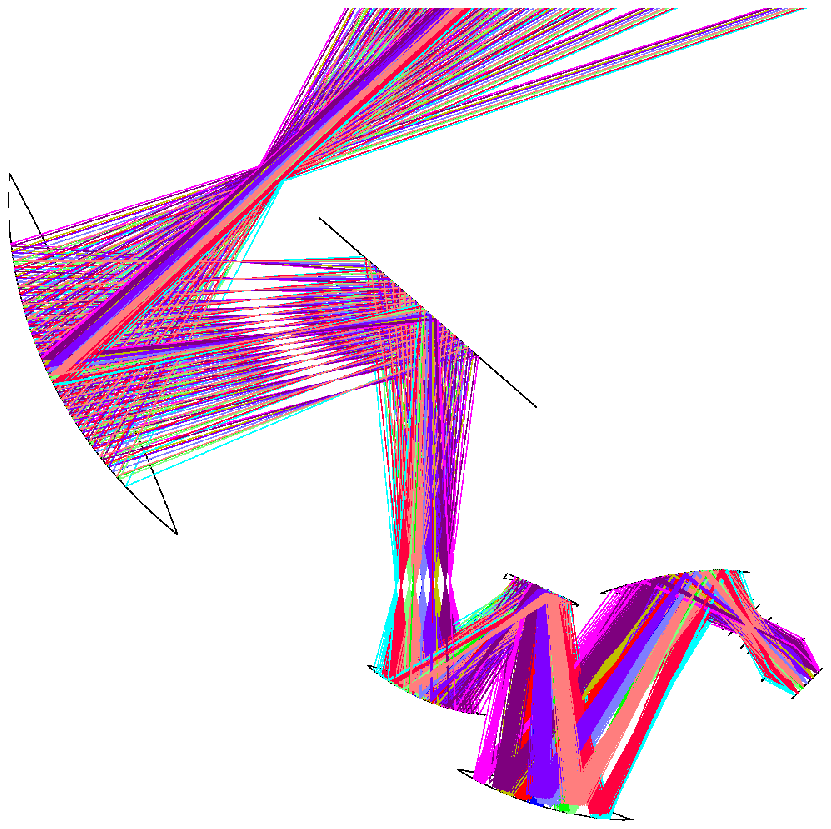}
\vskip 20 pt
\end{minipage}
\begin{minipage}{2.6in}
\caption[Detail view of Figure~\ref{fig:bigsideall}]
{{\bf Detail view of Figure~\ref{fig:bigsideall}.}
Rays from twelve points in the field are shown.
Note that rays from different points in the field cross 
each other at an
image of the aperture located between the
secondary and tertiary mirrors.
\label{fig:smallsideall}
\vfill}
\vskip 10 pt
\end{minipage}
\end{center}
\end{figure}

\section{The Problem of the Lagrange Invariant}

The design difficulties inherent in a wide-field system can be appreciated
by consideration of the Lagrange invariant.  One way of
expressing this quantity is 
${\mathit{Inv}}={{4}\over{\pi}}(A \Omega)^{1/2}$, where at a plane intersecting
the optical path, $A$ is the area illuminated
by the envelope of all rays through the system,
and $\Omega$ is the solid angle subtended by
the frustum of those rays.  This quantity is conserved along the
optical path.  Evaluating at the aperture of
the SPT telescope, 
${\mathit{Inv}} = 10 \mathrm{m} \cdot 1^{\circ} = 10\, {\mathrm{meter-degrees}}$.
If the beam passes through a small aperture some place in
the optical chain, for example at a 200 mm diameter dewar window, then
the beam there must converge at a fast $50^{\circ}$ angle.
It is impossible to pass the beam through a 50 mm diameter aperture,
because the included angle of the beam would need to be $200^{\circ}$.
The optical path will necessarily consist of large optical 
elements, of order a meter in diameter, and the beam will spread
rapidly between elements, requiring that the separation between
optical elements is comparable to their size. 

\begin{figure}[t!]
\begin{center}
\begin{minipage}{3.5in}
\vskip 10 pt
\epsfxsize=3.5in
\ \ \epsfbox{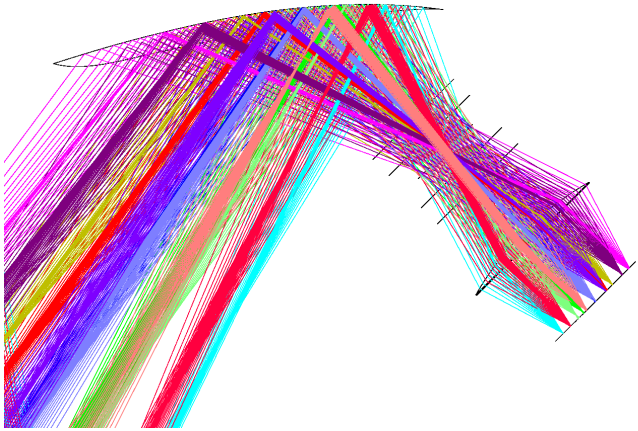}
\vskip 20 pt
\end{minipage}
\begin{minipage}{2.6in}
\caption[Detail view of Figure~\ref{fig:smallsideall}]
{{\bf Detail view of Figure~\ref{fig:smallsideall}.}
The last mirror and dewar optics are shown.
Note that the surface of best focus is flat, that the chief
ray from each point in the field is perpendicular to the image
plane, and that each of the points in the field uniformly illuminate
a Lyot stop near the dewar window.
The two lines in front of the Lyot stop, representing the
approximate position of the dewar window, and the two lines behind the
Lyot stop, representing the position of the IR blocking filters,
are 250 mm long and spaced 50 mm apart.
\label{fig:detectorall}
\vfill}
\vskip 10 pt
\end{minipage}
\end{center}
\end{figure}

Accommodating a large Lagrange invariant is awkward in 
conventional Cassegrain or Gregorian 
telescope designs, because the small size of the secondary
mirror and the large distance between it and its focus
make the subsequent optics gigantic.
The magnification of a telescope's secondary mirror, $m$, 
is the distance from the mirror center to
the far focus divided by the distance from the
mirror center to the near focus.
(By definition, the sign of $m$ is negative for Gregorians and positive
for Cassegrains.)
The secondary mirror magnification in on-axis Cassegrain or Gregorian
telescopes is typically $|m| \sim 25$.
Let $f_1$ is the focal length of the primary mirror.
In a 10 meter class telescope, $f_1 \sim 7\, \mathrm{m}$.
The size of the image at the Cassegrain focus is 
$m \cdot f_1 \cdot \mathrm{tan}(\mathit{FOV})$
where $\mathit{FOV}$ is the field-of-view.
If ${\mathit{FOV}} = 1^{\circ}$,
$m \cdot f_1 \cdot \mathrm{tan}({\mathit{FOV}})\sim 3 \, \mathrm{m}$,
so that any tertiary mirror in the vicinity of
the the Cassegrain or Gregorian focus must be about 3 meters across---an
awkwardly large size.
The way around this is to make $|m|$ small, closer to 3 than 25.

If $|m|$ is small, the Cassegrain or Gregorian focus is not
far from the secondary mirror, and the subsequent optics and
the detector will necessarily be located between the secondary
and the primary.  Such a position can easily be accommodated in an
off-axis telescope design, where the
detector dewar and its ancillary optics can be placed in the
secondary mirror support structure, out of the way of the primary
beam.  The National Radio Astronomy Observatory's Green Bank Telescope 
is configured this way.

\newpage

\begin{figure}[tb!]
\begin{center}
\begin{minipage}{3.7in}
\vskip 10 pt
\epsfxsize=3.7in
\ \ \epsfbox{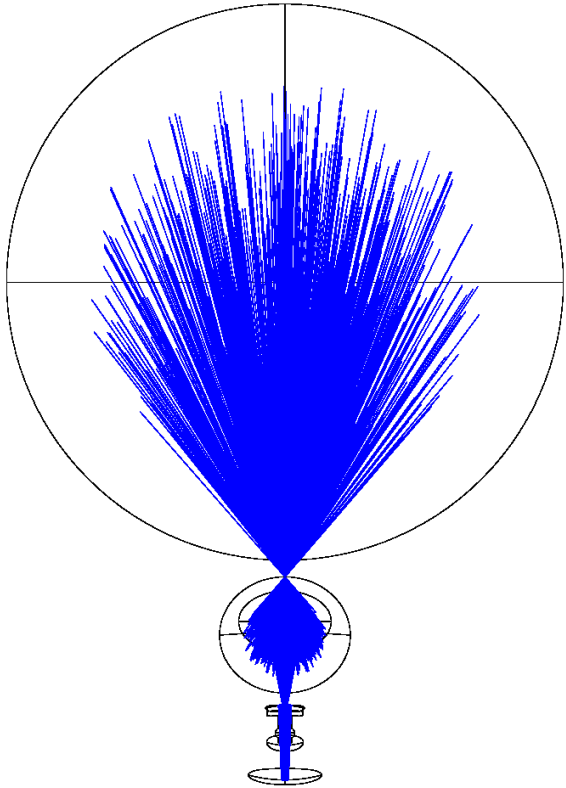}
\vskip 20 pt
\end{minipage}
\begin{minipage}{2.4in}
\caption[Star's view of SPST214]
{{\bf Star's view of SPST214\ \ }
A randomly-chosen but edge-tapered set of rays from the central 
point in the field of
view are shown.
Note that the optics are bilaterally symmetric to ensure that
horizontal chopping is balanced.
\label{fig:frontf1}
\vfill}
\vskip 10 pt
\end{minipage}
\end{center}
\end{figure}

\section{A Specific Design}

In optical design, it is difficult to prove that something
cannot be done---there are many clever tricks available to the
optical designer, and much room for the exercise of 
ingenuity.   It is, however, possible to show that something can
be done through an example.  
What follows is a particular optical design 
that meets the criteria for the Sunyaev-Zel'dovich 
effect survey with the South Pole
Telescope.  This is one possible design out of many.
It is not fully optimized and the South Pole
Telescope will not be built exactly this way, but it is presented
as a demonstration that a wide-field submillimeter-wave telescope is
possible.  
Other designs can be found at {\tt http://www.tonystark.org}.
The important point for the development of
Terahertz technology is the serendipitous result that this design 
exceeds the requirements for millimeter-wave work and 
is sufficiently well-corrected that it will
work at Terahertz frequencies.   

\renewcommand{\thefootnote}{\fnsymbol{footnote}}
The figures show various aspects of a ZEMAX\footnote{ZEMAX is a
trademark of ZEMAX Development Corporation of San Diego, CA} 
\renewcommand{\thefootnote}{\arabic{footnote}}
computer model 
of an optical design called SPST214.  
In this design, the primary mirror is a 10 m diameter section of 
of a paraboloid with a 7 m focal length.  The secondary is an
off-axis piece of a prolate ellipsoid.
Figure 1 is a view from the side, showing seven mirror surfaces, one
lens, and the detector plane.  Also shown are some rays from
a source at the center of the field of view: note that all
these rays come together at the prime focus, at the Gregorian
focus, and at the detector.  Figure 2 is the same view, but
shows rays from twelve sources distributed over the
one-degree-diameter field of view.  Figures 3, 4 and 5 are details
of Figures 1 and 2.  Figure 6 shows the view from the front,
as seen from the source being observed.  Note that the 
secondary mirror is well out of the main beam from the primary
mirror.  The secondary mirror is 2330 mm wide by 1970 mm high.
The tertiary mirror is flat, and will be used as a chopper.
Note in Figure 4 that it is located near an image of the
primary mirror.   The surfaces of mirrors four through seven 
have arbitrary shapes---their only constraint 
is that they are bi-laterally symmetric about a vertical
plane.  From the dewar window to the detector, the optical elements
are cylindrically symmetric about the chief ray of the center
of the field of view.

As seen in Figure 5, the dewar window is 250 mm diameter and will
likely be made of expanded plastic foam.  The Lyot stop is an
aperture 164 mm in diameter in the 4~K shield of the dewar. 
Infrared blocking filters are located behind the Lyot stop.
The diameter of these filters can be as small as the Lyot stop, or,
if they can be made larger, they could be located several centimeters
behind it, which would allow the Lyot stop to block some of the photons
scattered by the filter.

Silicon is an excellent material for a cold lens at millimeter wavelengths.
The lens in this design is 350 mm in diameter and about 25 mm thick.
A lens blank this size made of fairly high purity Silicon is readily
obtained from several sources.
The index of refraction of Silicon at 1.5 K is 3.38 (\cite{loewenstein73}),
and the loss through a lens this thick is
a few percent or less, depending on the purity of the material.  
The lens is constrained to be plano-convex in order to
facilitate anti-reflection coating.  A more ``bent'' lens would 
work a little better in this application, but the anti-reflection
coating might delaminate from a concave surface.
The lens is slightly aspheric.

\begin{figure}[tb!]
\begin{center}
\begin{minipage}{3.7in}
\begin{center}
\vskip 10 pt
\epsfxsize=1.7in
\ \ \epsfbox{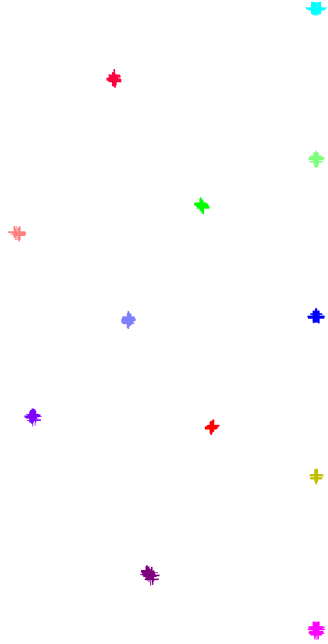}
\vskip 20 pt
\end{center}
\end{minipage}
\begin{minipage}{2.4in}
\caption[Spot diagram for SPST214]
{{\bf Spot diagram for SPST214\ \ }
This diagram shows the locations where various
rays from various stars in the field of view
intersect the image plane at 
the detector.  Only the left half of the
image plane is shown, because the telescope is
symmetric.  Each point of intersection is shown
with a cross.  Note that the rays from a given
star fall in a tight cluster.
The stars are distributed over a one degree
diameter field of view, which maps to a 200 mm 
diameter circle at the image.
\label{fig:spot}
\vfill}
\vskip 10 pt
\end{minipage}
\end{center}
\end{figure}

In the 150 GHz detector array to be used
in the Sunyaev-Zel'dovich survey, the converging cones of rays will
be intercepted by 1027 conical feedhorns drilled into an aluminum plate
which is 200 mm in diameter and about 8 mm thick.  There will
be a short length of waveguide between each feedhorn and each
bolometer detector, and that waveguide will encorporate 
a bandpass filter which will determine the
bandwidth of the observations.  The bolometers will be the
``spiderweb'' voltage-biased superconducting transition-edge
sensor type (\cite{gildemeister99}, 2000).

Figure 7 shows a spot diagram, where rays from twelve 
stars in a one degree diameter field of view are propagated
to the detector, and the intersection of each ray with the
detector plane is marked with a ``+''.
The important point for Terahertz technology is that the
quality of the image at the detector plane is good enough
to be used at Terahertz frequencies.  The root-mean-square
wavefront error at the detectors in this
design is typically $50\mu\mathrm{m}$.  This error could be
reduced a factor of $\sim 2$ by relaxing the quality of 
the image of the aperture at the Lyot stop, which 
in this design meets stringent requirements imposed by its
use as a microwave background anisotropy instrument.
A wavefront error of $50\mu\mathrm{m}$ is totally unsuitable for
use at visual wavelengths, which is why wide-field visual
wavelength telescopes are not designed this way,
but the design begins to be usable at 
$\lambda = 200 \, \mu \mathrm{m}$ and would work well at
$\lambda = 350 \, \mu \mathrm{m}$.  
Providing it could be made, an appropriate Terahertz
array could simply be substituted for the 150 GHz array 
currently being developed, and the optical system would work.
At Terahertz frequencies, the 200 mm diameter image plane
could accommodate tens of thousands of detectors.

\section{Conclusion}

The South Pole Telescope, which is nearing the end of its
design phase, will have an optical configuration capable of
feeding several $\times 10^4$ detectors at Terahertz frequencies.
Such a detector array would be enormously powerful.
Consider, for example, the problem of surveying an area of
sky to find dusty protogalaxies.  These are objects about one
second of arc across, with an areal density of several hundred
per square degree at a  brightness of a few mJy at
$\lambda350 \, \mu \mathrm{m}$.  The speed at which such
objects can be found is proportional to 
$${{n A B}\over{{T_{sys}^{*}}^2}} \, ,$$
where $n$ is the total number of detectors (bolometer or
heterodyne) in the telescope, $A$ is the total effective collecting
area (regardless of whether it is arranged in one dish
or many), $B$ is the pre-detection bandwidth, and
$T_{sys}^{*}$ is the effective system temperature adjusted for
losses, atmospheric absorption and emission (\cite{stark00}).

\begin{table}[ht]
\caption{Estimated Survey Speed at $\lambda350\, \mu \mathrm{m}$}
\begin{tabular} {|l|r|r|r|r|r|} \hline \hline
\emph{Telescope}  & $n$ & $A$ & $B$ & $T_{sys}^{*}$ & \emph{Relative Speed} \\ \hline
SPT & 20000 & $50\, \mathrm{m^2}$ & 150 GHz & 200 K & 3750 \\    
ALMA & 160 & $6400\, \mathrm{m^2}$ & 4 GHz & 300 K & 46  \\ \hline \hline  
\end{tabular}
\vskip 10 pt
\end{table}

\noindent
Table 1 estimates the value of this figure of
merit for the South Pole Telescope and for the
Atacama Large Millimeter Array (ALMA), and shows that the SPT with
a large bolometer array would be about 80 times faster than the
ALMA at detecting objects in survey fields.  Of course
the ALMA has much higher angular and frequency
resolution, and 
therefore shows a detailed picture. The SPT will be to
the ALMA what a Schmidt camera is to a large visual-wavelength
telescope---the means of detecting the objects to be studied.
All that is needed is the technological development of large
detector arrays.

\bigskip

I thank Nils Halverson, Adrian Lee, and William Holtzapfel
for suggesting the use of a Silicon lens in the design.
This material is based upon work supported by the National
Science Foundation under Grant No. OPP-0130612.


\begin{thebibliography}

\bibitem[{Gildemeister, Lee, \& Richards 1999}]{gildemeister99}
Gildemeister, J., Lee, A., \& Richards, P. 1999, Appl. Phys. Lett., 74, 868

\bibitem[{Gildemeister, Lee, \& Richards 2000}]{gildemeister00}
---. 2000, Appl. Phys. Lett., 77, 4040

\bibitem[{Loewenstein, Smith, \& Morgan 1973}]{loewenstein73}
Loewenstein, E.~V., Smith, D.~R., \& Morgan, R.~L. 1973, Appl. Opt., 12, 398

\bibitem[{Stark 2000}]{stark00}
Stark, A.~A. 2000, in Proceedings of SPIE, Vol. 4015, Radio Telescopes, ed.
  H.~R. Butcher, 434

\bibitem[{Stark {et~al.} 1998}]{stark98b}
Stark, A.~A., Carlstrom, J.~E., Israel, F.~P., Menten, K.~M., Peterson, J.~B.,
  Phillips, T.~G., Sironi, G., \& Walker, C.~K. 1998, in Advanced Technology
  MMW, Radio and Terahertz Telescopes, Vol. 3357 (Proceedings of SPIE),
  495--506, {\tt astro-ph/9802326}

\end{thebibliography}
\end{document}